\documentclass[intlimits,twoside,a4paper]{article}

\usepackage[cp1251]{inputenc}
\usepackage[eqsecnum]{cmpj3}

\issue{2023}{26}{1}{13503}
\doinumber{10.5488/CMP.26.13503}

\title{Merging of Dirac points through uniaxial modulation on an optical lattice}

\author[A. L\'opez, B. Monta\~nes, E. Medina]{A. L\'opez\orcid{0000-0002-1220-6036}\refaddr{label1}, B. Monta\~nes\orcid{0000-0001-8047-8546}\refaddr{label2},
	E. Medina\orcid{0000-0002-1566-0170}\refaddr{label3}}
\addresses{
	\addr{label1}Escuela Superior Polit\'ecnica del Litoral, ESPOL, Departamento de F\'isica, Facultad de Ciencias Naturales y Matem\'aticas, Campus Gustavo Galindo Km. 30.5 Via Perimetral, P. O. Box 09-01-5863, Guayaquil, Ecuador
	\addr{label2}Centro de F\'{\i}sica, Instituto Venezolano de
	Investigaciones Cient\'{\i}ficas, Caracas, Venezuela
	\addr{label3}Departamento de F\'isica, Colegio de Ciencias e Ingenier\'ia, Universidad San
	Francisco de Quito, Diego de Robles y V\'ia Interoce\'anica, Quito, 170901,
	Ecuador}

\Keywords{cold gases in optical lattices, graphene, topological phase transition, Bloch-Floquet theorem}

\date{Received October 04, 2022, in final form January 04, 2023}

\begin{document}

	\maketitle

\begin{abstract}
We analyze the scenario of modulating the potential strength of bound atoms in an optical honeycomb lattice patterned by an electric field to emulate uniaxial strain. This modulation can be achieved by a combination of the strength of the patterned electric field and gauge vector effects using the Floquet approach. We show that such a modulation allows one to follow through a topological transition between a semi-metal and a band insulator, when two non-equivalent $K$ points merge as a function of the electric field strength. We explicitly compute the wavefunctions for the moving $K$ points and the Chern numbers up to the transition. Anisotropic effective masses and the insulating gap are described close to the semimetal-insulator transition.

\printkeywords

\end{abstract}

\section{Introduction}

Twistronics is the most recent field to generate surprises in condensed matter physics~\cite{MacDonald}. As it turns out in bilayer graphene, with a relatively straightforward single electron physics, twisting one layer with respect to the other produces a modulation in the bands such that non-trivial phases appear that involve strong electron correlations~\cite{MacDonald}. The generic scenario seems to be that of band flattening producing heavy fermions and making electron-electron interactions a prime ingredient. This way, non-trivial superconducting and insulating phases arise in experiments~\cite{Cao,Park}. This transition is achieved by fine tuning bilayer interactions by twisting to particular angles. Another role of deformation in graphene, is the relative strength of its spin-orbit coupling (SOC), central in its topological properties. While ideal flat graphene has a SOC in the \textmu{}eV range, bending the sheet into nanotubes increases it three orders of magnitude into the meV range~\cite{Huertas}. The hydrogenation of graphene also produces local deformations that enhance $\pi$--$\sigma$ coupling and thus the SOC~\cite{FabianHydroGraphene}. 

In a realization of $SU(2)$ gauge theories in deformed graphene, Vozmediano, Katsnelson and Guinea~\cite{Vozmediano} have shown how to phrase the lattice deformations into gauge fields. Such gauge fields can be associated to spin-orbit magnetic and electric fields, that result in weak localization and Berry phase effects, where the particular structure of the 2D lattice sets the stage to unify concepts from elasticity and topology. 

Finally, in the context of quasi-one dimensional systems, e.g., molecular electronics, the SOC also arises from appropriate curvatures of the molecular system. The SOC has been proven to be the spin-active coupling through deformation effects of hydrogen bond polarization~\cite{HydrogenBondCISS,HydrogenBondCISS_2}.

In this work we consider a recent proposal of an optical lattice with the structure of graphene~\cite{Zhu,Loon} containing Alkaline atoms e.g., $^{40}$K with two mobile fermions per unit cell (half filling). They analyze the band structure with simple uniaxial deformations of the optical lattice by way of patterned electric fields of attainable magnitudes. Within this model one can follow the excursion of the $K$ points as a function of the electric field modulation, and as the Fermi level remains fixed, we can follow the Hamiltonian spectrum and eigenfunctions as a function of the modulation strength. We can then explicitly compute the topological Chern number for small deviations from the $K$ points and show how the topological protection disappears when the next order is included or we approach the merging of the $K$ points. The latter is also illustrated by the appearance of backscattering due to trigonal warping which is also modulated by deformation. Finally, we consider additional weaker electric fields, generating phase effects on the fermion motion. We can then see with finer detail the nature of the topological transition from Dirac cones to parabolic bands.  

We dedicate this work to Bertrand Berche on his 60th birthday. Bertrand, has been a close collaborator for more than 16 years and much of our most impactful work was done with his collaboration, advice, and his knack for mastering new mathematical tools to address exciting problems outside his traditional field of expertise.
\section{Modulation of bonding by patterned electric field}\label{section2}
Optical lattices have been used for long time~\cite{CohenTannoudji} to establish a $d$-dimensional binding potential that organizes atoms into crystals in e.g., a Bose condensate~\cite{ReviewOpticalLattices}. For multilevel atoms, to achived the appropriate potential, one uses appropriate laser detuning $\delta=\omega_L-\omega_{at}$, where $\omega_L$ is the laser frequency and $\omega_{at}$ are the frequencies of the energy level splittings of the atom. If $\delta$ is much larger than any of the atomic transitions, all such transitions will experience the same dipole potential of the form~\cite{Loon}
\begin{equation}
    V({\bf r})=\frac{\hbar\Gamma}{8}\frac{\Gamma}{\delta}\frac{I({\bf r})}{I_s},
\end{equation}
where $I({\bf r})=\varepsilon_0 c|E({\bf r})|^2/2$ and $I_s$ is the saturation intensity of the particular atom. $\Gamma$ is the angular frequency width of the excited state in a particular transition, which in the high detuning limit will have a common value. An interesting scenario is to introduce an appropriate superposition of electric field phases and orientations to concoct the desired lattice potential. Following the formulation of optical lattices in ref.[~\cite{Loon}], we choose an electric field perpendicular to the resulting lattice plane of the form
\begin{equation}
    {\bf {\cal E}}_a=E_0\, \re^{\ri({\bf k}_l\cdot {\bf r}-\phi_a)}\re^{-\ri\omega_L t}{\hat z},
\end{equation}
where $E_0$ is the electric field intensity and the ${\bf{k}}_l$ vectors with $l=1,2,3$ satisfy ${\bf k}_1+{\bf{k}}_2+{\bf{k}}_3=0$. In order to build a potential of a honeycomb lattice with primitive vectors ${\bf{a}}_1=a/2(1,\sqrt{3})$ and ${\bf{a}}_2=a/2(1,-\sqrt{3})$ and reciprocal vectors ${\bf{b}}_1=2\piup/a(1,1/\sqrt{3})$ and ${\bf{b}}_2=2\piup/a(1,-1/\sqrt{3})$, we need the superposition of the following ${\bf{k}}_l$ vectors  
\begin{eqnarray}\label{eq3}
    {\bf k}_1&=&{\bf K}=\left(\frac{4\piup}{3a}\xi,0\right)=\frac{ {\bf b}_1+{\bf b}_2}{3},\nonumber\\
    {\bf k}_2&=&{\bf K}-{\bf b}_1,\nonumber\\
    {\bf k}_3&=&{\bf K}-{\bf b}_2,
\end{eqnarray}
satisfying the null sum, where ${\bf K}$ is the $K$ point vector of the Brillouin zone and $\xi=\pm 1$. 

All parameters of the electric field are chosen so that the energy minima of the optical potential fall at the very sites of the graphene lattice. The coupling between these sites is now modulated by the optical potential, whose phases $\phi_l$ can be chosen so that differences arise in the nearest neighbor coupling in a chosen way that we specify in the next section. 

This pattern of electric fields generates minima when the light is blue detuned from the atomic resonance and $\delta>0$. The fermion bearing atoms are dropped in the potential and assume the minima, thus organizing into a honeycomb lattice, with identical potential peaks separating A and B sublattice atoms. For $^{40}$K alkali atoms, this differs from graphene in that we have an $s$-wave model, but addressing strict two dimensional physics, like in-plane distortions, there is not a distinction to the $p$-wave physics of material graphene.

\section{Graphene uniaxial modulation}\label{section3}
Graphene~\cite{geim0,geim1,geim2,Kane2005} is a two-dimensional material composed of carbon atoms distributed in a honeycomb structure. Here, we are interested in studying fermions on a honeycomb lattice when we induce uniaxial~\cite{arias} modulation of the dipole potential. As we have proposed, modulating the crystal potential by a patterned electric field we equate to stretching and compressing the `bond', meaning weakening/strengthening the overlap integral between two sites. To study the electronic properties of this model, we use the tight-binding approach. By means of such approximation, we have the Hamiltonian given by 

\begin{equation}\label{eq4}
    H({\bf k})=\left(\begin{array}{cc}
        0 & f({\bf k}) \\
        f^*({\bf k}) & 0
    \end{array}\right),
\end{equation}
where $f({\bf k})$ has dimensions of energy according to
\begin{equation}
    f({\bf k})=-\sum_{l=1}^{3}t_l \re^{\ri\bf{k}\cdot {\delta_l}},
\end{equation}
with $t_l$ the $s$ orbital overlaps (in plane rotational symmetric) to neighbor $l$
and $\delta_1 =(0,1/\sqrt{3})a$, $\delta_2=(1/2,-1/2\sqrt{3})a$, $\delta_3 =(-1/2,-1/2\sqrt{3})a$,
where $a$ is the A--A or B--B distance. With these vectors, the function $f({\bf k})$ becomes
\begin{equation}
    f({\bf k})=-t_1 \exp \left({\ri\frac{k_y a}{\sqrt{3}}}\right)-t_2 \exp \left[{\ri\frac{a}{2\sqrt{3}} \left(\sqrt{3}k_x-k_y\right)}\right]-t_3 \exp \left[{-\ri\frac{a}{2\sqrt{3}} \left(\sqrt{3}k_x+k_y\right)}\right].
\end{equation}
We determined the corresponding energy by solving the secular equation $\det|H-E\mathbf{I}|=0$, which gives the energy dispersion $E({\bf k})=\pm\sqrt{f({\bf k}) f^*({\bf k})}$. In turn, the wave function is
\begin{equation}
    \Psi=\frac{1}{\sqrt{2}}\left(\begin{array}{c}
         1  \\
         \xi \frac{|f({\bf k})|}{f({\bf k})}
    \end{array} \right) \re^{\ri {\bf k}\cdot{\bf r}}\ ,\ \ \ \ \ \xi=\pm 1.
\end{equation}
We define uniaxial stretching in the coordinate system chosen above as the change in the $t_1$ vector keeping $t_{2,3}$ fixed. This makes, for a particularly simple closed form, the evolution of the band parameters. We then set $t_1=t_1$ and $t_2 = t_3 = t_0$. In this case, we have

\begin{equation}
    f({\bf k})= -t_0\left[ \frac{t_1}{t_0} \exp \left({\ri \frac{k_y a}{\sqrt{3}}}\right)+2 \exp \left({-\ri\frac{k_y a}{2\sqrt{3}}}\right)\cos\left(\frac{k_x a}{2}\right)\right].
    \label{fofkuniaxial}
\end{equation}
Note, this modulation is difficult to achieve on actual graphene with stretching patterns, but straightforward on an optical lattice.

The energy dispersion is then
\begin{equation}
    \frac{E_{\pm}}{t_0}=\pm\sqrt{2+\left(\frac{t_1}{t_0}\right)^2+2\cos\left(a k_x\right)+4 \left(\frac{t_1}{t_0}\right) \cos\left(\frac{a k_x}{2}\right) \cos\left(\frac{\sqrt{3} }{2} a k_y\right)}.
    \label{EnergyStretched}
\end{equation}
For convenience, we define dimensionless parameters $\tilde E=E/t_0$, $\tilde f({\bf k})=f({\bf k})/t_0$ and $ t=t_1/t_0$ throughout.

\begin{figure}[!h]
	\centering
\includegraphics[width=5.5cm]{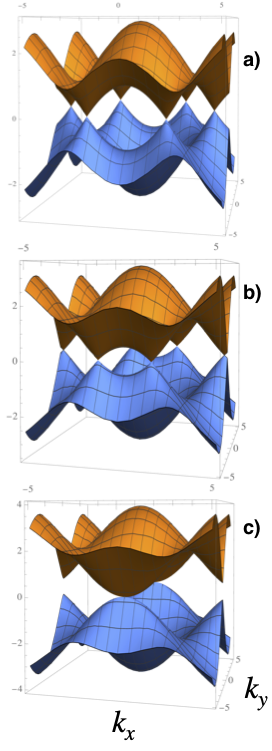}
\caption{(Colour online) The energy dispersion for different values of $1<t<2$. Note, Fermi energy is unchanged under deformation and the Dirac cone physics is preserved up to the topological transition. The merging of the $K$ points occurs at ${\bf G/2}=\pm{({\bf b}_1-{\bf b}_2)}/{2}$.}
\label{LinearDispersiont>1}
\end{figure}

\section{The Dirac Hamiltonian with uniaxial deformation}
In this section we study the behavior of fermions on the honeycomb optical lattice (for brevity we call it optical graphene) under uniaxial modulation. For this, we follow the band properties near the Dirac points. The Dirac points as a function of $t$ can be found by setting $f({\bf K_{\xi}})=0$ in equation~\eqref{fofkuniaxial} and solving for $K_{\xi}$. For $t=1$, ${\bf K}_{\xi}=(4\piup/3a) {\hat x}$, while for $1<t<2$
\begin{equation}
    { {\bf K}_{\xi}}=\xi \left(\frac{2}{a}\arccos\left(-\frac{ t}{2}\right),0\right),\ \ \ \ \ \xi=\pm 1.
\end{equation}
We define ${\bf q}={\hbar \bf
 k}-{\hbar \bf K}_{\xi}$, then for small $|{\bf q}|$ to lowest order, we find
\begin{equation}
    {\tilde f}({\bf k})=\frac{a}{2\hbar}\left({ \xi\sqrt{4- t^2}} q_x-\ri{\sqrt{3}} t q_y\right).
\end{equation}
Writing $q$ in polar coordinates: ${\bf q}=q(\cos\varphi, \sin\varphi)$ the effective Hamiltonian is given by
\begin{equation}
     {\tilde H}=\frac{aq}{2\hbar}\left(\sqrt{4- t^2}\xi \cos\varphi~\sigma_x+\sqrt{3} t\sin\varphi ~\sigma_y\right)
\end{equation} 
and the eigenvalue is
\begin{equation}
    {\tilde E}=\pm \frac{a q}{2\hbar}\sqrt{ t^2-2(t^2-1)\cos2\varphi+2}\, ,
\end{equation} 
which displays an obvious anisotropy with respect to the momentum direction. The Berry phase around the $K$ point, labelled by $\xi=\pm 1$, is computed as~\cite{Hasegawa2006,Montambaux,Fuchs2010}
\begin{equation}
    \gamma= \frac{1}{2}\oint\nabla_{\bf q}\theta_{\bf q}\cdot \rd{\bf q},
\end{equation}
where $\theta_{\bf q}=\tan^{-1}[\Im f({\bf q})/\Re f({\bf q})]$. In the linear approximation gives $\xi \piup$, independent of the value of~$t$. Of course, this is an artifact of the linear approximation and the threshold for the topological transition is evident in the next order in the deviation from the $K$ points. 
 
\begin{figure}[!t]
	\centering
\includegraphics[width=13cm]{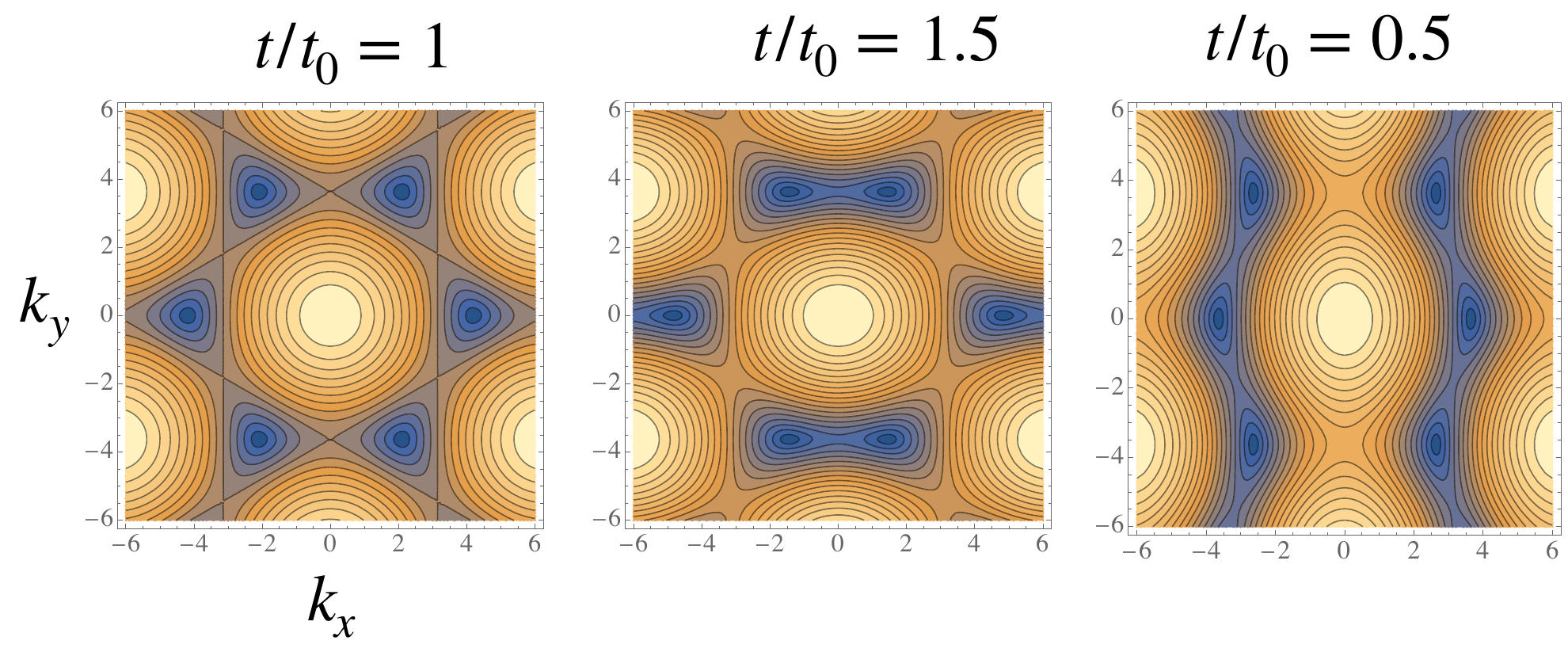}
\caption{(Colour online) Dispersion in the full BZ as a function of deformation. When $\tilde{t}\rightarrow 0$, the system breaks up into uncoupled chains (right-hand panel). Such a limit is achieved by the merging of $K$~points into a line. In the limit $\tilde{t}\rightarrow 2$, the system converges to a normal insulator through the topological transition (middle panel).}
\label{Contours0<t<2}
\end{figure}

When one goes beyond the lowest order, we find
\begin{eqnarray}
    {\tilde f}({\bf k})&=& \frac{a}{2\hbar}\left(\xi \sqrt{4-t^2} q_x-\ri\sqrt{3}t q_y\right)\nonumber\\
    &-& \frac{a^2}{8\hbar^2}\left(t q_x^2+2 \ri\xi\sqrt{\frac{4-t^2}{3}}q_x q_y -t q_y^2\right).
    \label{f(k)}
\end{eqnarray}
Thus, using the polar form for $q$ one obtains that the Hamiltonian reads
\begin{eqnarray}
         H&=& \left(\xi\frac{\sqrt{4-t^2}}{2\hbar}a q\cos\varphi -\frac{t a^2q^2}{8\hbar^2}\cos2\varphi\right) \sigma_x \nonumber \\
         &+&\frac{\sqrt{3}}{6\hbar}a q \sin\varphi\left(3 t+\xi\frac{\sqrt{4-t^2}}{2\hbar}a q\cos\varphi\right) \sigma_y.
\end{eqnarray} 
The corresponding eigenvalue is then
\begin{align}
   & {\tilde E}=\pm \frac{a q\sqrt{3}}{24} \nonumber \\
    \times& \sqrt{\left(48+a^2 q^2\right) \left(2+t^2\right)-96 \left(t^2-1\right) \cos2 \varphi+2 a q \left[-12 b t \sqrt{4-t^2} \cos3\varphi +a q \left(t^2-1\right) \cos4\varphi\right]}.
\end{align}
The wavefunction in its general form is
\begin{equation}
    \Psi=\frac{1}{\sqrt{2}}\left(\begin{array}{c}
         1  \\
         \xi \frac{|f({\bf k})|}{f({\bf k})}
    \end{array} \right) \re^{\ri {\bf k}\cdot{\bf r}}
\end{equation}
and can be written to the previous approximation.
We depict the energy dispersion at the linear approximation in figure~\ref{LinearDispersiont>1}, where we can see that uniaxial deformation produces two lobes that pinch off as the two $K$ points merge. As depicted in figure~\ref{Contours0<t<2}, we see the linear approximation and the effect of higher order terms in the limit $t=1$. The angular dependence is known as trigonal warping~\cite{mccan} of the Dirac cones as we go away from the Fermi energy. 

The interest of following the Berry phase is that it yields a threshold for $t$ at which it becomes zero. This signals the occurrence of a topological transition that coincides with the merging of two Dirac cones.
In the linear approximation, the Berry phase is computed as [see equation~\eqref{f(k)}]
\begin{equation}
    \theta_{\bf p}=\tan^{-1}\left(\frac{-\sqrt{3}p_y}{\sqrt{4-{t}^2}\xi p_x}\right).
\end{equation}
By changing to polar coordinates around the $K({t})$ point, for $0<{t}<2$, we substitute $p_x=p \cos\varphi$ and $p_y=p \sin\varphi$ and we get
\begin{equation}
    \theta_{\bf p}=\tan^{-1}\left(\frac{-\sqrt{3}}{\xi\sqrt{4-{t}^2}}\tan\varphi\right).
\end{equation}
${\nabla}=\hat{\bf p}~\partial/\partial p+\hat{\varphi} (1/p)\partial/\partial \varphi$ and $\rd{\bf p}=\hat{\bf p}\rd p+p\hat{\varphi}\rd\varphi$ so we arrive at the expression for the Berry phase as
\begin{eqnarray}
    \gamma({t})&=&-\frac{\xi\sqrt{3}}{2}\int_0^{2\piup}\frac{\sqrt{4-{t}^2}\sec^2\varphi}{4-{t}^2+3\tan^2\varphi} \rd\varphi\nonumber\\
    &=&-\xi\piup,
\end{eqnarray}
for all values of $0<t<2$. Thus, there is a wide range of $\tilde{t}$ values where there is a single topological phase untouched by modulations of the uniaxial coupling.


\section{Backscattering amplitude}
The scattering amplitude as a function of the momentum direction has a special significance in graphene showing a backscattering protection related to the topological properties. 

\begin{figure}[h]
	\centering	
	\includegraphics[scale=0.3]{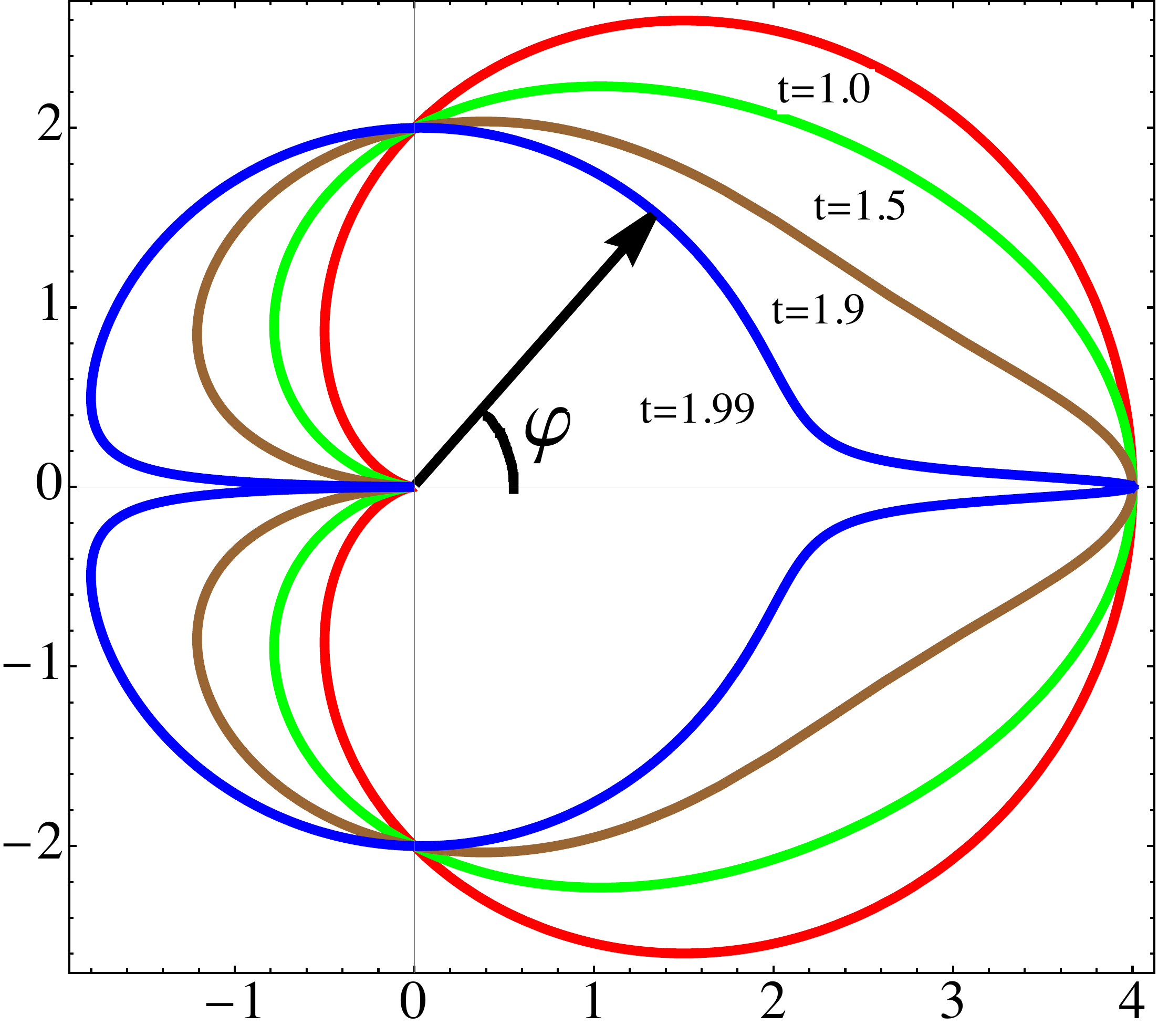}
	\caption{(Colour online) The unnormalized probability of scattering at an angle of $\varphi$, $|\langle \Psi(\varphi)|\Psi(0)\rangle|^2$, for uniaxial compression. In consistency with topological protection, backscattering is forbidden before the topological transition.}
	\label{Scatt-t>1}
\end{figure}

\begin{figure}[]
	\centering
	\includegraphics[scale=0.3]{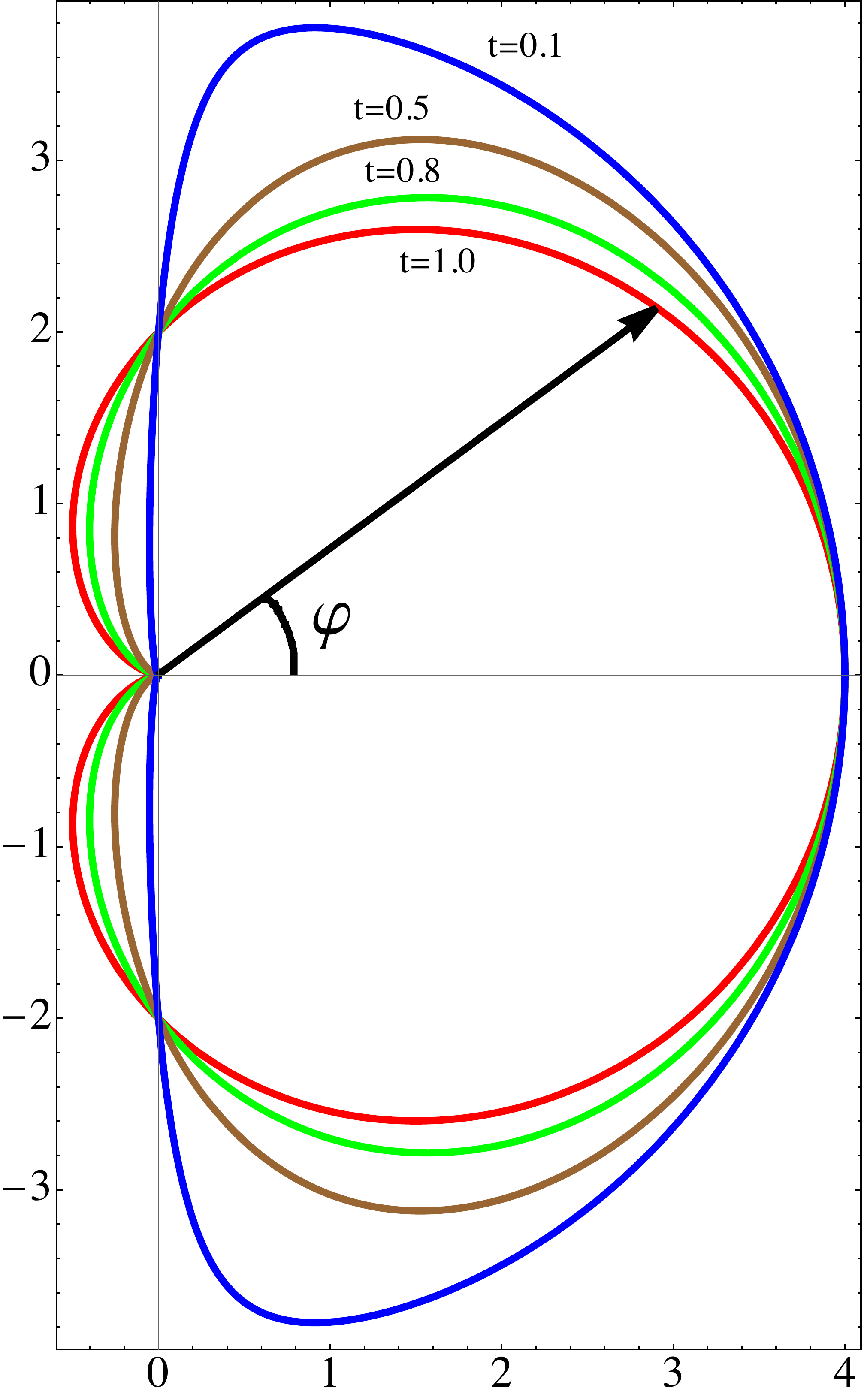}
	\caption{(Colour online) The unnormalized probability of scattering at an angle of $\varphi$, $|\langle \Psi(\varphi)|\Psi(0)\rangle|^2$ for uniaxial stretching.}
	\label{Scatt-t<1}
\end{figure}

Figures \ref{Scatt-t>1} and \ref{Scatt-t<1} show $|\langle \Psi(\varphi)|\Psi(0)\rangle|^2$ in a polar plot. The undeformed case concentrates scattering in the forward direction while forbidding $\varphi=\piup$ or back scattering as a signature of topological protection. Such signature is preserved in the range of uniaxial modulation $0<t<2$. For $t\rightarrow 0$ shows the tendency to form isolated chains with a pronounced forward scattering while preserving topological protection. When $t=0$, we have chains with a half sized primitive cell, as A--B become equivalent, and we arrive at a one dimensional dispersion. For $t\rightarrow 2$, we get a highly peaked forward scattering amplitude when the two Dirac cones approach each other. Topological protection disappears at $t=2$ with a quadratic dispersion in the $x$ direction while keeping a linear dispersion (massless) in the $y$ direction. We see there that the scattering occurs equally in all directions.

\section{Effective mass}
For deformations between $0<t<2$, the optical graphene Hamiltonian only exhibits massless Weyl fermions close to the $K$ points. It is only when we have two merging Dirac points that a non-relativistic effective mass arises. This occurs at the point $\xi(2\piup/a,0)$ on the edge of the Brillouin zone. As the merging $K$ points have opposite topological charges, initially there is zero total topological charge and after the merger this is conserved.

We can obtain the effective masses for each direction by approaching the merging point from close to $t>2$ starting from the general relation for the energy equation~\eqref{EnergyStretched}. Taking $t=2+\delta t$ with $\delta t>0$
and ${\bf k}={(q_x+2\piup/a,q_y)}$ we obtain the expression
\begin{equation}
    \frac{E}{t_0}=\pm\left(|t-2|+\frac{3a^2 t q_y^2}{4|t-2|}+\frac{a^2 q_x^2 ~{\rm sgn}(t-2)}{4
    }+ \cdots \right),
\end{equation}
where we can identify the gap as $\Delta=2 (t-2)$ vanishing at the transition and the effective masses $m^*_x=2\hbar^2~{\rm sgn (t-2)}/a^2$ and $m^*_y=2\hbar^2|t-2|/3a^2 t$. As can be verified by an expansion below the topological transition at $t=2$, we have a relativistic dispersion in the $y$ direction, so the classical effective mass vanishes and $m^*_x$  does not depend on deformation close to the transition. 

\section{Driven scenario}\label{sec7}

Besides the dipolar field due to the patterned electric field, we can consider another linearly polarized electric field in another frequency and power regime to modulate the A--B coupling in optical-graphene. This way we have another route to drive the system through the topological transition, now involving both the dipole field and the gauge field. 
Taking advantage of this combination, in this section we explore a minimalist model for the dynamical realization of Dirac cone merging, by means of electromagnetic radiation. To this end, we consider a monochromatic laser field, linearly polarized, which impinges perpendicularly to the graphene plane. Therefore, within the tight binding formulation, using the Peierls substitution ${\bf k}\rightarrow {\bf k}+e{\bf A}(\tau)$, with $\tau$ being the time parameter, $-e$ is the electron's charge, and ${\bf A}(\tau)$ gives the vector potential, describing the electromagnetic radiation. 
We write the time-dependent Hamiltonian from \eqref{eq4}, which now reads
\begin{equation}
    H({\bf k}, \tau)=\left(\begin{array}{cc}
        0 & f({\bf k},\tau) \\
        f^*({\bf k},\tau) & 0
    \end{array}\right),
\end{equation}
where we have defined
\begin{equation}
    f({\bf k},\tau)=-\sum_{l=1}^{3}t_l \exp\left\{{\ri[{\bf k}+e{\bf A}(\tau)]\cdot \delta_l}\right\}.
\end{equation}
To describe the anisotropic nature of the Dirac cone merging phenomena discussed within the static scenario, we focus our analysis on the dynamical modulation of the system by means of a linearly polarized radiation field, which without loss of generality, can be chosen along the $x$ direction, i.e.,  ${\bf E}(\tau)=E(\cos\Omega \tau,0)$, where $\tau$ represents the time parameter, whereas $E$ and $\Omega$ are the amplitude and frequency of the radiation field, respectively. Then, using the relation ${\bf E}=-\partial_\tau {\bf A}(\tau)$, the vector potential is found to read ${\bf A}(\tau)=E/\Omega(\sin\Omega \tau,0)$. In the limit $\Omega\rightarrow 0$, we recover the static scenario previously described and we can have a clear physical description of the role of the driving field in producing the topological transition.  

The generic driven scenario for periodically driven systems is treated within the Floquet formalism. Although the Floquet theory has appeared more often in the literature, mostly thanks to the theoretical proposals and experimental realizations of Floquet topological insulators (FTI)~\cite{NP2011Lindner}, which are the nonequilibrium counterparts of topological insulators~\cite{VonKlitzing1980,Haldane1988,Bernevig2006,Koenig2007,Hasan2010,Bernevig2014,3DTI}, we briefly give an account of the main ingredients emerging within the driven regime (further details can be found in reference~\cite{LopezVarela}). From the time-dependent Schr\"odinger equation, the periodic Hamiltonian $H(\tau)=H(\tau+T)$ generates the time evolution $\ri\hbar\partial_\tau\Psi(\tau)=H(\tau)\Psi(\tau)$.  The solution can be written as $\Psi(\tau)=\exp({\ri\epsilon \tau/\hbar})\Phi(\tau)$, with $\Phi(\tau+T)=\Phi(\tau)$ being the periodic part of the solution, which is termed the Floquet state and $\epsilon$ is the quasi energy~\cite{Grifoni1998,chu}. 

The quasi energy spectrum stems from the fact that the Hamiltonian is time-dependent, implying that the energy is not a conserved quantity and it is similar to the quasi momentum in spatially periodic systems. In addition, given the periodic nature of the dynamical generator, the quasienergies are also periodic since they are defined modulo $\hbar\Omega$. In general, the explicit evaluation of the quasi energy spectrum can be achieved by Fourier expanding the Hamiltonian and Floquet states to obtain a time-independent set of equations for the time independent Floquet Hamiltonian $H_{\rm F}=P^\dagger(\tau)H(\tau)P(\tau)-\ri\hbar P^\dagger(\tau)\partial_\tau P(\tau)$. This Floquet Hamiltonian generates the time-independent Schr\"odinger equation for the Floquet states defined previously: $H_{\rm F}\Phi(\tau)=\epsilon\Phi(\tau)$.

In order to generate the Dirac cone merging given in reference~\cite{Montambaux}, within the driven scenario we can obtain an effective Hamiltonian with differential hopping parameters along two directions in the lattice, i.e., one requires breaking the rotational invariance of the static Hamiltonian. We can achieve this condition by means of the linearly polarized radiation field, which permits the manipulation of the hopping strength along the direction of the driving field. Moreover, to get a physical insight we focus on the so-called high frequency regime, where $\hbar\Omega$ is much larger than the other energy scales in the problem. Within this regime, we obtain an analytically solvable anisotropic quasi energy spectrum.

To this end, we write the explicit time-dependence of $f({\bf k},\tau)$ as~\cite{LopezVarela}    
\begin{equation}
    f({\bf k},\tau)=-\left[t_1+t_2 \re^{\ri\xi\sin\Omega\tau}\re^{{\ri\bf k}\cdot {\bf a}_1}+t_2\re^{-\ri\xi\sin\Omega\tau}\re^{{\ri\bf k}\cdot {\bf a}_2}\right].
\end{equation}
Here, we have defined the effective dimensionless light-matter coupling strength $\xi={\sqrt{3}eEa}/{2\hbar\Omega}$.

The phase factor can be rewriten by means of the Jacobi-anger expansion: 
$$\re^{\ri\xi\sin\Omega\tau}=\sum_{-\infty}^{\infty} J_n(\xi) \re^{\ri n\Omega\tau},$$
where $J_n(x)$ is a Bessel function of the first kind of order $n$.
In the long frequency regime, all but the zero order terms oscillate very quickly and therefore, one gets $\exp({\ri\xi\sin\Omega\tau})\rightarrow J_0(\xi)$, with $J_0(x)$ representing the zeroth-order Bessel function of the first kind. Whence, if we define $t_0=t_2J_0(\xi)$, we get the effective expression
\begin{equation}
    f({\bf k})=-\left[t_1+t_0 \left(\re^{{\ri\bf k}\cdot {\bf a}_1}+\re^{{\ri\bf k}\cdot {\bf a}_2}\right)\right].
\end{equation}
Upon the addition of a static contribution $\Delta\sigma_x$ we can realize the topological phase transition proposed in reference~\cite{Montambaux}. Within this realm, the evaluation of the quasi-energy spectrum for the high frequency regime considered here gives $\varepsilon_\pm=\pm \varepsilon$, with
\begin{equation}
\varepsilon=\sqrt{(t_1+\Delta)^2+4t_0\cos \left(\frac{\sqrt{3}k_xa}{2}\right)\left[t_1\cos\left(\frac{3k_ya}{2}\right)+t_0\cos\left(\frac{\sqrt{3}k_xa}{2}\right)\right]}.
\end{equation}
The corresponding Floquet states are given by
\begin{equation}
|\phi_\mu(k)\rangle=\frac{1}{\sqrt{2}}
\begin{pmatrix}
1\\
\mu \re^{\ri\beta}
\end{pmatrix},
\end{equation}
with $\mu=\pm1$ and the phase is defined through
\begin{equation}
\tan\beta=\frac{\Im{f({\bf k})}}{\Re{f({\bf k})}}.  
\end{equation}
The condition for realizing the Dirac cones merging is $t_1=2t_0$ (i.e., $t=2$ in the dimensionless notation defined before), which in turn implies that the radiation field parameter $\xi=\xi_c$ should satisfy $J_0(\xi_c)=1/2\rightarrow \xi_c\approx1.5$. This in turn would correspond to an effective electric field intensity of the order $E=\sqrt{3}\hbar\Omega/ea$. Some theoretical works have set $\hbar\Omega=3t_1$, with $t_1=2.9$~eV being graphene's hopping parameter; yet, even if we set $\hbar\Omega=t_1$ and using the carbon-carbon spacing $a=0.142$~nm, the condition for realizing the driven Dirac cone merging topological phase would require an electric field intensity of order $E\approx35$~V/nm. This result is one order of magnitude higher than the experimental electric field intensities $E=2.4$~V/nm employed in reference~\cite{higuchi}, where light-induced currents are analyzed in graphene. However, photonic alternatives such as the microwave tight binding analogue model, presented in reference~\cite{MontambauxPRL2013}, where the hopping energy can be properly tuned, could serve as test bed for the experimental realization of our proposal.  

\begin{figure}
    \includegraphics[height=5.5cm]{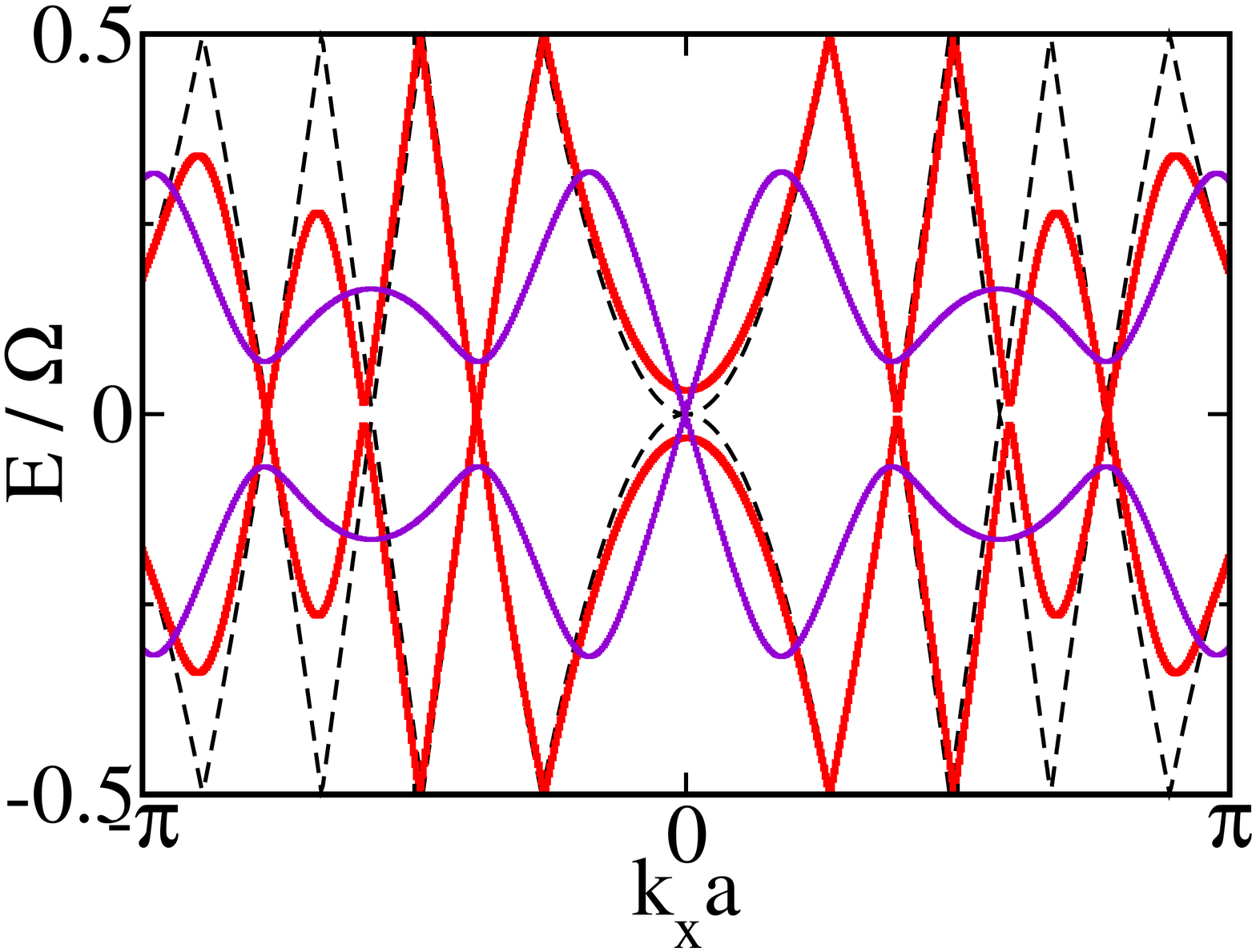}
    \includegraphics[height=5.5cm]{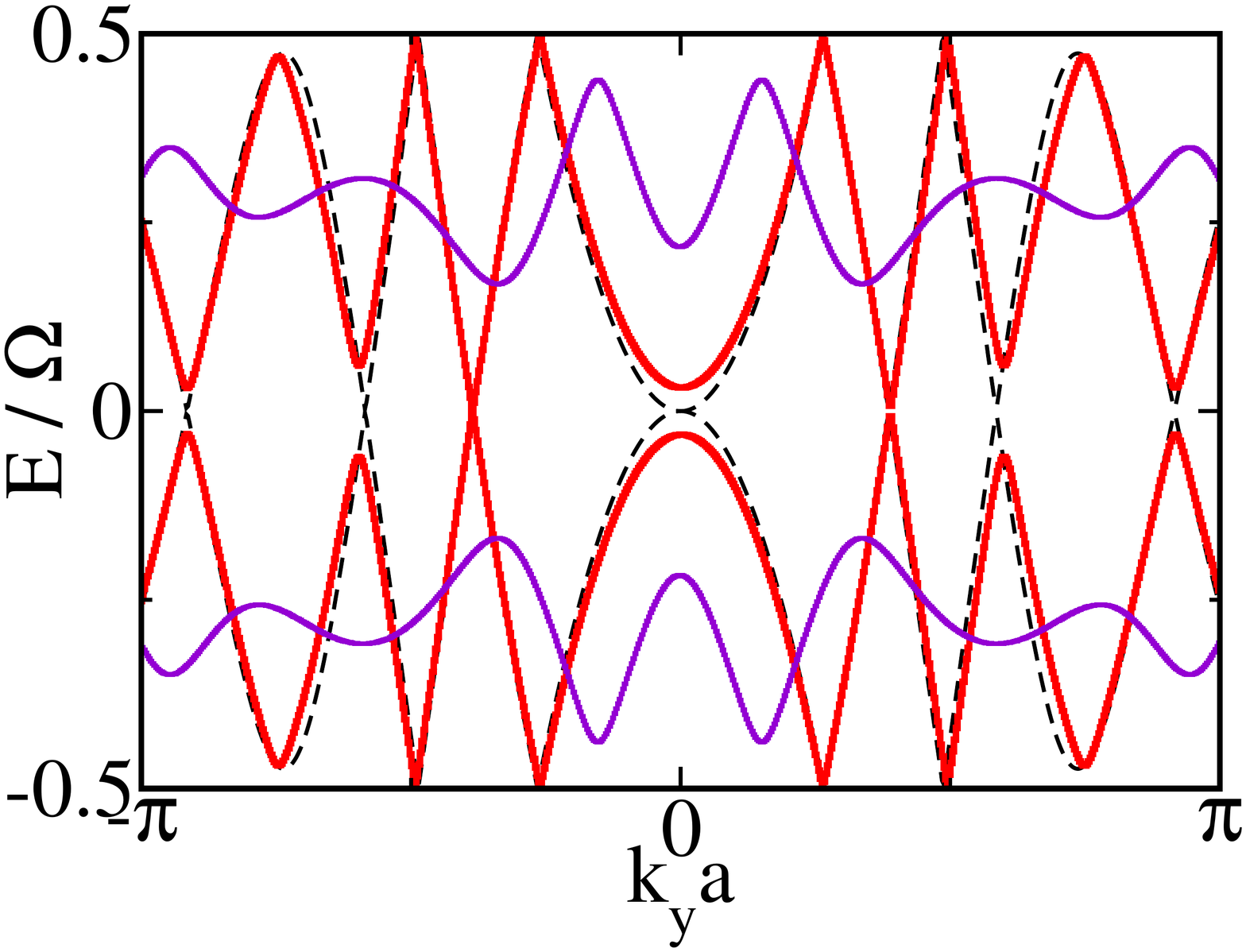}
    \caption{(Colour online) The dashed lines correspond to the static scenario, the red (thick) lines correspond to the low coupling regime $\xi=0.25$ whereas the magenta (thin)  lines give the strong $\xi=1$ coupling regime, at low frequencies. We have set the bias parameter $\Delta=0$ (see the main text).}
    \label{fig:my_label}
\end{figure}

At low frequencies, the higher-order Floquet replica become relevant and the evaluation of the quasi energy spectrum requires a numerical evaluation of the  Floquet Hamiltonian. For reference of the results obtained at low frequencies, we show the resulting quasinergy spectrum for $n=30$, harmonics for a frequency value smaller than the static bandwidth of pristine graphene. The left-hand (right-hand) panel in figure~\ref{fig:my_label} shows the quasi energy spectrum along the applied (perpendicular) direction of the radiation field. The dashed thin lines correspond to the static spectrum which is wrapped in the first Brillouin zone $-\Omega/2<E<\Omega/2$. The red (thick) and magenta (thin) continuous lines represent the low (high) coupling regimes to the radiation field. In the low coupling regime, the effective dimensionless light-matter interaction is set to $\xi=0.25$ and we already observe qualitative differences corresponding to the anisotropic generation of bandgaps at finite values of momenta. In the strong coupling regime, at zero momentum, the direction along the applied radiation field becomes gapless whereas the perpendicular direction remains a gap and this is reminiscent of the previously discussed high frequency regime where the anisotropy of the energy spectrum shows linear and parabolic spectra along the chosen momentum directions. In order to highlight the role of the radiation field in inducing the bandgap modulation, we have chosen a static bias term $\Delta=0$.

\section{Summary and conclusions}
Following the topological transition from a semi-metal and a band insulator in pristine graphene is physically unfeasible through the modulation of C--C bonding since it would require large changes of the overlap integrals~\cite{Montambaux}. On the other hand, honeycomb lattices can be achieved with cold atoms and optical lattices~\cite{Loon} with reasonable patterned electric
fields creating dipole potentials where one can engineer potential anisotropies and tune them to the transition and beyond. Furthermore, with fields in a very different frequency range and through gauge effects, one can also modulate the $^{40}$K--$^{40}$K potentials and finely walk across the transition. Although this setup builds an artificial s-wave graphene model, the in-plane physics is indistinguishable and pristine one electron Dirac-cone physics ensue with negligible electron-electron and spin-orbit interactions.

Uniaxial modulation of the $^{40}$K--$^{40}$K potentials can be followed analytically and the Berry-phase can be computed exactly for all coupling values from zero anisotropy to the uniaxial strain producing the transition. We explicitly compute the consequences of the uniaxial modulation on the topological protection from backscattering and compute the effective masses approaching the transition from above (uniaxial strains beyond the merging of the $K$ points), showing non-zero effective masses in one direction and massless dispersion in the perpendicular direction.

\section*{Acknowledgements}
We dearly thank Bertrand for more than 16 years of exciting collaborations in the most diverse topics, technical and pedagogical, that have been well received in the Condensed Matter and Physical Chemistry literature in very influential journals. Congratulations Bertrand on your 60th birthday!



\newpage

\ukrainianpart

\title{Злиття точок Дірака через одновісну модуляцію на оптичній ґратці}
\author{А. Лопес\refaddr{label1}, Б. Монтаньєс\refaddr{label2}, Е. Медінa\refaddr{label3}}

\addresses{
	\addr{label1}Вища політехнічна школа Побережжя, факультет фізики, природничих наук та математики, Кампус Густаво Галіндо
	30.5 км Віа Періметрал, 09-01-5863, Гуаякіль, Еквадор
		\addr{label2}Фізичний центр Венесуельського інституту природничих досліджень, Каракас, Венесуела
		\addr{label3} Фізичний факультет, Науково-технічний коледж, Університет Святого Франциска в Кіто, Дієго-Робелес і Віа Інтерокеаніка, Кіто, 170901, Еквадор}
\makeukrtitle 

\begin{abstract}
	\tolerance=3000%
	Проаналізовано сценарій модуляції параметра інтенсивності зв’язаних атомів на оптичній стільниковій ґратці, створеній електричним полем для імітації одновісної деформації. Ця модуляція може бути досягнута шляхом поєднання напруженості структурованого електричного поля та калібрувальних векторних ефектів за допомогою підходу Флоке.
	Ми показуємо, що така модуляція дозволяє прослідкувати топологічний перехід між напівметалом та зонним ізолятором, коли дві нееквівалентні $K$-точки зливаються в залежності від напруженості електричного поля. Явним чином розраховано хвильові функції для рухомих $K$-точок та чисел Черна аж до самого переходу. Анізотропні ефективні маси та щілина в спектрі описані поблизу переходу ``напівметал-ізолятор''.
	
	\keywords{холодні гази на оптичних ґратках, графен, топологічні фазові переходи, теорема Блоха-Флоке}
	
\end{abstract}

\lastpage

\begin{thebibliography}{66}
\bibitem{MacDonald} MacDonald A. H., Physics, 2019, \textbf{12}, 12, \doi{10.1103/Physics.12.12}.
\bibitem{Cao}Cao Y., Fatemi V., Fang S., Watanabe K., Taniguchi T., Kaxiras E., Jarillo-Herrero P., Nature, 2018, {\bf 556}, 43,\\ \doi{10.1038/nature26160}.
\bibitem{Park} Park J. M., Cao Y., Watanabe K., Taniguchi T., Jarillo-Herrero P., Nature, 2021, {\bf 590}, 249, \doi{10.1038/s41586-021-03192-0}.
\bibitem{Huertas} Huertas-Hernando D., Guinea F., Brataas A., Phys. Rev. B, 2006, {\bf 74}, 155426, \doi{10.1103/PhysRevB.74.155426}.
\bibitem{FabianHydroGraphene} Gmitra M., Kochan D., Fabian J., Phys. Rev. Lett., 2013, {\bf 110}, 246602, \doi{10.1103/PhysRevLett.110.246602}.
\bibitem{Vozmediano} Vozmediano M. A. H., Katsnelson M. I., Guinea F., Phys. Rep., 2010, {\bf 496}, 109,\\ \doi{10.1016/j.physrep.2010.07.003}.
\bibitem{HydrogenBondCISS} Torres J. D., Hidalgo-Sacoto  R., Varela S., Medina E., Phys. Rev. B, 2020, {\bf 102}, 035426,\\ \doi{10.1103/PhysRevB.102.035426}.
\bibitem{HydrogenBondCISS_2} Salazar S. V., Mujica V., Medina E., Chimia, 2018, {\bf 72}, 411, \doi{10.2533/chimia.2018.411}.
\bibitem{Zhu}Zhu S.~L., Wang B., Duan L.~M., Phys. Rev. Lett., 2007, {\bf 98}, 260402, \doi{10.1103/PhysRevLett.98.260402}.
\bibitem{Loon}Lee K. L., Gr\'emaud B., Han R., Englert  B.~G., Miniatura C., Phys. Rev. A, 2009, {\bf 80}, 043411,\\ \doi{10.1103/PhysRevA.80.043411}.
\bibitem{CohenTannoudji}Verkerk P., Lounis B., Salomon C., Cohen-Tannoudji C., Courtois J.~Y., Grynberg G., Phys. Rev. Lett., 1992, {\bf 68}, 3861, \doi{10.1103/PhysRevLett.68.3861}.
\bibitem{ReviewOpticalLattices} Grynberg G., Robilliard C., Phys. Rep., 2001, {\bf 355}, 335, \doi{10.1016/S0370-1573(01)00017-5}.
\bibitem{geim0} Novoselov K. S., Geim A. K., Morozov S. V., Jiang D., Zhang Y., Dubonos S. V., Grigorieva I. V., Firsov A. A., Science, 2004, {\bf 306}, 666, \doi{10.1126/science.1102896}.
\bibitem{geim1} Geim A. K., Novoselov K. S., Nat. Mater., 2007, {\bf 6}, 183, \doi{10.1038/nmat1849}.
\bibitem{geim2} Castro Neto A. H., Guinea F., Peres N. M. R., Novoselov K. S.,  Geim A. K., Rev. Mod. Phys., 2009, {\bf 81}, 109,\\ \doi{10.1103/RevModPhys.81.109}.
\bibitem{Kane2005}Kane C. L., Mele E. J., Phys. Rev. Lett., 2005, {\bf 95}, 226801, \doi{10.1103/PhysRevLett.95.226801}.
\bibitem{arias} Arias E., Hern\'andez A. R., Lewenkopf C., Phys. Rev. B, 2015, \textbf{92}, 245110, \doi{10.1103/PhysRevB.92.245110}.
\bibitem{Hasegawa2006}Hasegawa Y., Konno R., Nakano H., Kohmoto M., Phys. Rev. B, 2006, \textbf{74}, 033413,\\ \doi{10.1103/PhysRevB.74.033413}.
\bibitem{Montambaux}Montambaux G., Pi\'echon F., Fuchs J.~N., Goerbig M. O., Phys. Rev. B, 2009, \textbf{80}, 153412,\\ \doi{10.1103/PhysRevB.80.153412}.
\bibitem{Fuchs2010}Fuchs J.~N., Pi\'echon F., Goerbig M. O., Montambaux G., Eur. Phys. J. B, 2010, \textbf{77}, 351, \doi{10.1140/epjb/e2010-00259-2}.
\bibitem{mccan} McCann E., In: Graphene Nanoelectronics. NanoScience and Technology, Raza H. (Ed.), Springer, Berlin, Heidelberg, 2012, 237--275, \doi{10.1007/978-3-642-22984-8\_8}.
\bibitem{NP2011Lindner} Lindner N. H., Refael G., Galitski V., Nat. Phys., 2011, {\bf 7}, 490, \doi{10.1038/nphys1926}.
\bibitem{VonKlitzing1980}Klitzing  K. V., Dorda G., Pepper M., Phys. Rev. Lett., 1980, {\bf 45}, 494, \doi{10.1103/PhysRevLett.45.494}.
\bibitem{Haldane1988}Haldane  F. D. M., Phys. Rev. Lett., 1988, {\bf 61}, 2015, \doi{10.1103/PhysRevLett.61.2015}.
\bibitem{Bernevig2006} Bernevig B. A., Hughes T. L., Zhang S.~C., Science, 2006, {\bf 314}, 1757, \doi{10.1126/science.1133734}.
\bibitem{Koenig2007}K{\"o}nig M., Wiedmann S., Br{\"u}ne C., Roth A., Buhmann H., Molenkamp  L. W., Qi X.~L., Zhang S.~C., Science, 2007, {\bf 318}, 766, \doi{10.1126/science.1148047}.
\bibitem{Hasan2010}Hasan M. Z., Kane C. L., Rev. Mod. Phys., 2010, {\bf 82}, 3045, \doi{10.1103/RevModPhys.82.3045}.
\bibitem{Bernevig2014}Alexandradinata A., Fang C., Gilbert M. J., Bernevig B. A., Phys. Rev. Lett., 2014, {\bf 113}, 116403,\\ \doi{10.1103/PhysRevLett.113.116403}.
\bibitem{3DTI}Xu Y., Miotkowski I., Liu C., Tian J., Nam H., Alidoust N., Hu J., Shih C.~K., Hasan M. Z., Chen Y. P., Nat.~Phys., 2014, {\bf 10}, 956, \doi{10.1038/nphys3140}.
\bibitem{LopezVarela} L\'opez A., Varela S., Medina E., J. Phys.: Condens. Matter, 2022, {\bf 34}, 135301, \doi{10.1088/1361-648x/ac48c1}.
\bibitem{Grifoni1998} Grifoni M., H\"anggi P., Phys. Rep., 1998, {\bf 304}, 229, \doi{10.1016/S0370-1573(98)00022-2}.
\bibitem{chu} Chu S.~I., Telnov D. A., Phys. Rep., 2004, {\bf 390}, 1, \doi{10.1016/j.physrep.2003.10.001}.
\bibitem{higuchi}Higuchi T., Heide C., Ullmann K., Weber H. B., Hommelhoff P., Nature, 2017, {\bf 550}, 224,\\ \doi{10.1038/nature23900}.
\bibitem{MontambauxPRL2013}Bellec M., Kuhl U., Montambaux G., Mortessagne F., Phys. Rev. Lett., 2013, {\bf 110}, 033902,\\ \doi{10.1103/PhysRevLett.110.033902}.


\end{thebibliography}
\end{document}